# RECONSTRUCTION OF BEAM TRANSVERSE PARAMETERS IN THE FERMILAB SIDE-COUPLED LINAC USING A NORMALIZED COORDINATE FRAMEWORK

E. V. Chen†, J-P. Carneiro, R. V. Sharankova, A. V. Shemyakin
Fermi National Accelerator Laboratory, Batavia, IL, USA

*Abstract*

Quadrupole scans are a commonly used tool for beam second moment reconstruction. Limitations in the strength of the magnets and layout of the beamline elements frequently preclude simple quadrupole-drift-detector scans from collecting sufficient data for reconstruction. Using a normalized coordinate framework, we characterize the prerequisites for a robust simple quadrupole scan and expand these prerequisites to reconstruction from more complex optics. The beam second moments are investigated at two locations in the Fermilab Side-Coupled Linac under simple and complex optics, using this framework to maximize information gained from wire scanner profile measurements.

## INTRODUCTION

In the transverse planes, the quadrupole scan is the classical method for measuring beam second moments [1]. The simplest form of the quadrupole scan varies the current in a quadrupole and measures the beam size at a downstream detector, with only a drift between the two elements. Reconstruction of beam second moments occurs at the upstream quadrupole position. However, limitations in the quadrupole strengths, element layout, or permissible beam losses frequently prohibit successful 'simple' quadrupole scan measurements.

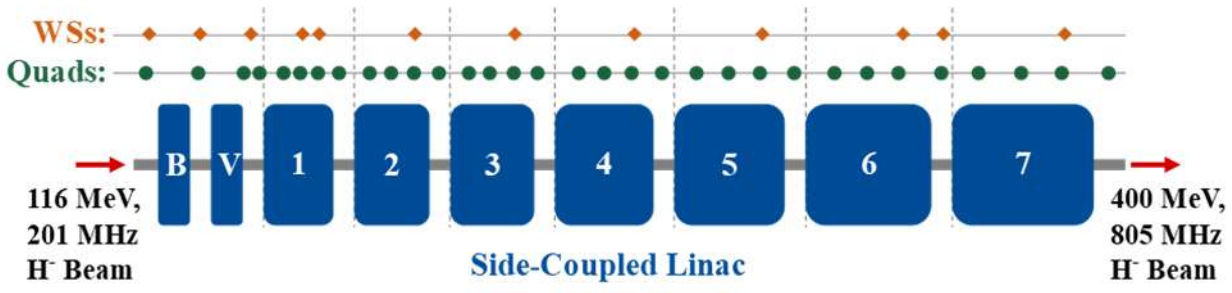

Figure 1: Side-Coupled Linac schematic.

The 'simple' quadrupole scan can be generalized to be compatible with any transfer matrix between the reconstruction point and detector(s). This freedom allows for mitigation of the 'simple' quadrupole scan limitations; however, more complex beam transport precludes simple, visual assessment of the quality of the reconstruction.

Here we use the Fermilab Side Coupled Linac (SCL) [2], Fig. 1, to conduct three measurements of beam second moments at different locations with the simple and generalized quadrupole scan techniques. We compare results from the reconstructions, assess the robustness of these setups via a normalized coordinate framework, and use this information to develop suggestions for more efficient measurements. Note that the space charge is mostly neglected in this work. Although beam second moment reconstruction only informs on rms properties of the beam, these principles similarly underly reconstruction techniques beyond the first order, such as in phase space tomography.

### *Quadrupole Scan Formulae*

The beam second moments, $\boldsymbol{\Sigma}$, are directly related to the Twiss parameters and emittance and are propagated via the transfer matrix, $\boldsymbol{M}$, from the initial upstream location, $i$, to the final downstream location, $f$, as shown in Eq. (1).

$$\boldsymbol{\Sigma}_f = \boldsymbol{M}\boldsymbol{\Sigma}_i\boldsymbol{M}^T = \begin{pmatrix}\sigma_{11,f} & \sigma_{12,f}\\ \sigma_{21,f} & \sigma_{22,f}\end{pmatrix} = \epsilon_f\begin{pmatrix}\beta_f & -\alpha_f\\ -\alpha_f & \gamma_f\end{pmatrix} \quad (1)$$

The $\sigma_{11}$ term of $\boldsymbol{\Sigma}$ is equal to the rms beam size squared, $\sigma_{rms}^2$. In the simple quadrupole scan with a thin lens approximation and ignoring space charge, the relationship between $\sigma_{11,f}$ and quadrupole strength, $kl$, is parabolic.

To generalize, the beam size $\sqrt{\sigma_{11,f}}$ is measured under varying transfer matrices between the upstream reconstruction point and downstream detector. The three unique second moments, $\sigma_{11,i}$, $\sigma_{12,i}$, and $\sigma_{22,i}$, can thus be solved for via Eq. (2) and (3), which reshape the relationship in Eq. (1) to isolate the collected beam sizes in $\boldsymbol{\eta}$ and leave the transfer matrix terms generalized but reshaped in $\boldsymbol{A}$ [3].

$$\begin{pmatrix}\sigma_{11,f}\\ \vdots\end{pmatrix} = \begin{pmatrix}m_{11}^2 & 2m_{11}m_{12} & m_{12}^2\\ \vdots & \vdots & \vdots\end{pmatrix}\begin{pmatrix}\sigma_{11,i}\\ \sigma_{12,i}\\ \sigma_{22,i}\end{pmatrix} \quad (2)$$

$$\boldsymbol{\eta} = \boldsymbol{A}\boldsymbol{\sigma} \quad (3)$$

The pseudoinverse of $\boldsymbol{A}$, $\boldsymbol{A}^+$, can be used to solve this simple minimization problem for $\boldsymbol{\sigma}$.

### *Normalized Coordinates*

In $(x, x')$ phase space, shear and expansion during beam transport complicates visualizing the optimal set of downstream projections for reconstruction of beam second moments. This visualization is simpler in normalized coordinates $(x_n, x_n')$ where the beam transport is rotation of a rigid body.

The transfer matrix can be decomposed into the betatron amplitude matrix, $\boldsymbol{\beta}$, and rotational matrix, $\boldsymbol{R}$, Eq. (4) [4].

$$\boldsymbol{M} = \boldsymbol{\beta}_f\boldsymbol{R}\boldsymbol{\beta}_i^{-1} \quad (4)$$

The transition to normalized coordinates $(x_n, x_n')$ occurs via incorporating the betatron amplitude matrices into the coordinates $(x, x')$, leaving behind only the rotational matrix, Eq. (5) and (6). The rotational angle is the betatron phase advance, $\psi$, and units are m$^{-1/2}$ for both $x_n$ and $x_n'$.

$$\begin{pmatrix}x_{n,f}\\ x_{n,f}'\end{pmatrix} = \begin{pmatrix}\cos\psi & \sin\psi\\ -\sin\psi & \cos\psi\end{pmatrix}\begin{pmatrix}x_{n,i}\\ x_{n,i}'\end{pmatrix} \quad (5)$$

$$x_n = \frac{x}{\sqrt{\beta}},\ x_n' = \frac{x\alpha}{\sqrt{\beta}} + x\sqrt{\beta} \quad (6)$$

† erinchen@fnal.gov

Transport from the upstream reconstruction point to the downstream detector is a pure rotation of the phase space object under normalized coordinates. If the normalized beam phase space distribution is close to a 2D Gaussian, the optimal distribution of sampling angles is expected to be evenly spaced in $(x_n, x'_n)$ [5].

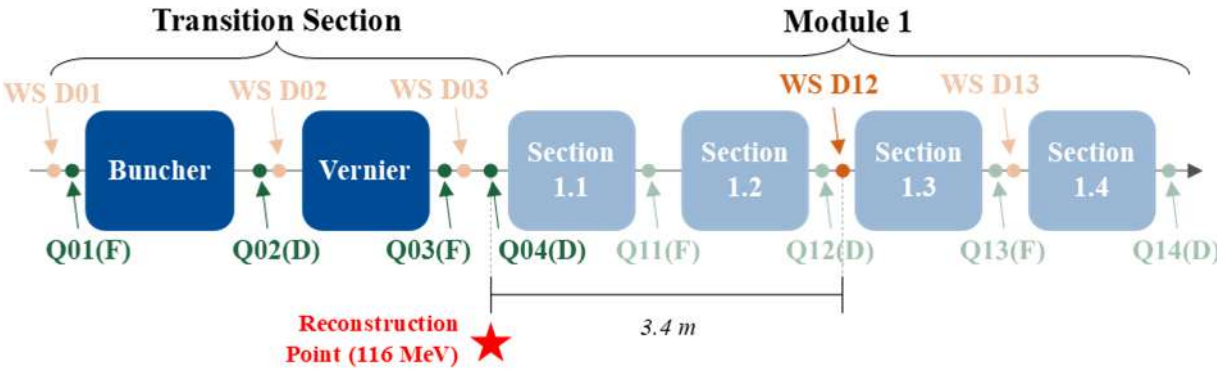

Figure 2: Simple quadrupole scans measurement design.

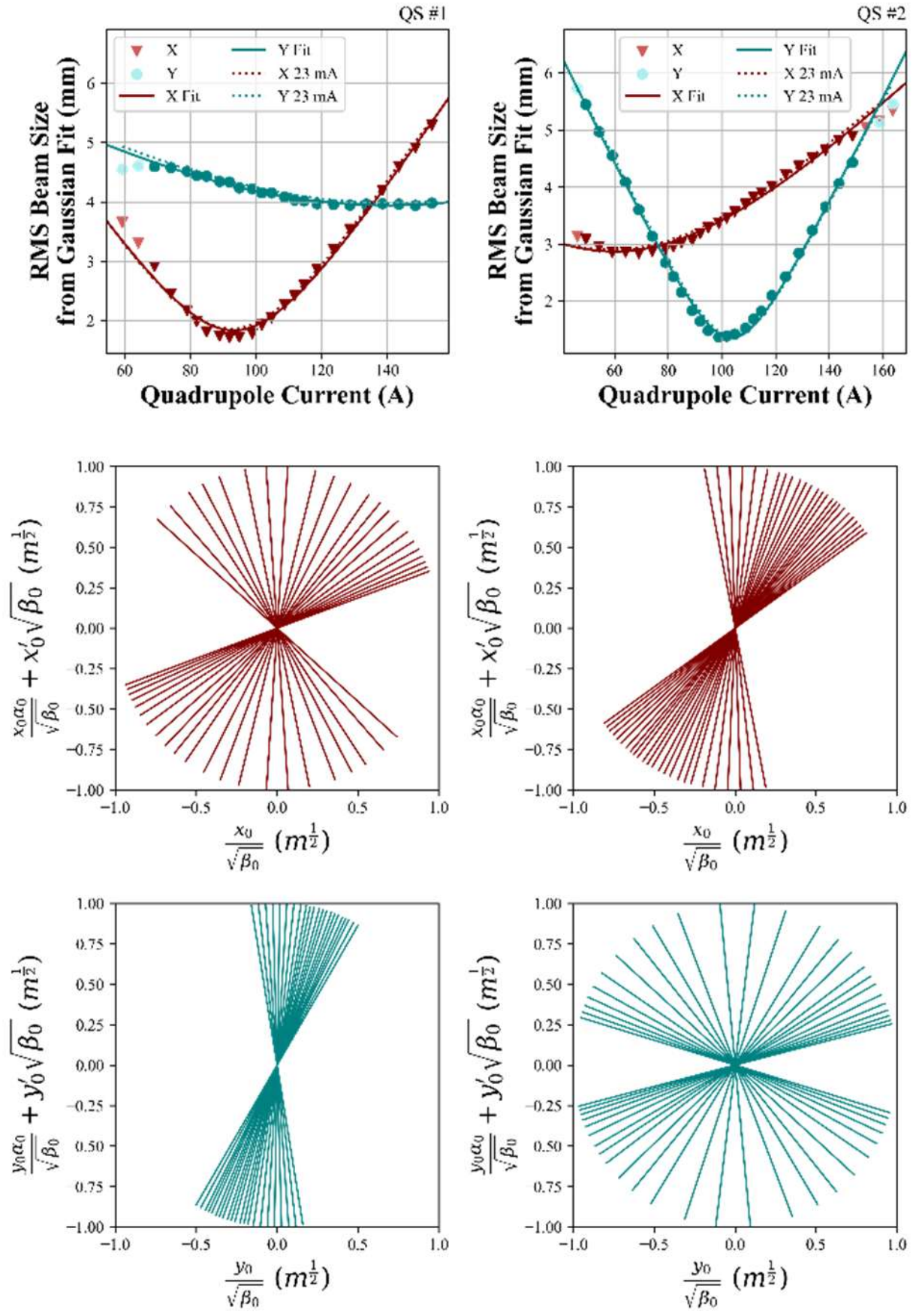

Figure 3: QS #1 (left) and QS #2 (right) results. Measured data, fits, and simulation with space charge (top), and phase advance in X (middle) and Y (bottom) under normalized coordinates.

## SIMPLE QUADRUPOLE SCAN

The FODO lattice of the SCL and single power supply powering the four quadrupoles in a module limit simple quadrupole-drift-detector scans to using the defocusing quadrupole at the end of a module (Qn4, where n = Module number) as the driving quadrupole. To create the drift, the accelerating cavities and quadrupoles in the n+1 Module are turned off. A wire scanner (WS) in the n+1 Module is used to collect beam transverse profiles, where the rms beam size is extracted from a Gaussian fit of the data.

Here, we collect simple quadrupole scans with Q04 and WS D12 under two sets of upstream optics: "QS #1" with adjustments to the two upstream bunching modules and the three upstream quads; and "QS #2" under nominal beamline conditions, Fig. 2. The scans are shown in Fig. 3 and the reconstructed Twiss parameters and normalized rms emittance are detailed in Table 1. Simulations at 23 mA using the reconstructed Twiss parameters show the effects of space charge are negligible, Fig. 3.

Table 1: Twiss Parameters and Error ($\sigma$)

| | QS #1 | QS #2 | QS #3 |
|---|---|---|---|
| $\epsilon_x$ (mm-mrad) | 0.81 ± 0.01 | 0.71 ± 0.03 | 0.86 ± 0.01 |
| $\beta_x$ (m) | 5.9 ± 0.2 | 2.06 ± 0.06 | 8.9 ± 0.2 |
| $\alpha_x$ | 6.0 ± 0.2 | 1.60 ± 0.09 | -3.7 ± 0.1 |
| $\epsilon_y$ (mm-mrad) | 0.77 ± 0.03 | 0.73 ± 0.02 | 0.79 ± 0.03 |
| $\beta_y$ (m) | 1.04 ± 0.04 | 8.5 ± 0.3 | 0.86 ± 0.03 |
| $\alpha_y$ | -0.86 ± 0.02 | -4.7 ± 0.1 | 0.49 ± 0.03 |

The vertical emittance agrees within 1σ, however the horizontal emittance agree within 3σ. (The upstream conditions of the two scans differed, and $\beta$ and $\alpha$ are not directly comparable.) The shown errors are statistical. Errors in second moments are estimated by the covariance matrix of the fit, $\boldsymbol{S}$, using the residual sum of squares $RSS$ and the degrees of freedom $d$, Eq. (7) [6]. These errors are propagated to the error in the emittance and Twiss parameters via the Jacobian as described in [7].

$$\boldsymbol{S} = (\boldsymbol{A}^T\boldsymbol{A})^{-1}\frac{RSS}{d} \tag{7}$$

We did not find a rigorous explanation for the discrepancy between the scatter in measured emittances and the low values of the assigned errors and speculate that it can be related to incomplete characterization of the beam waist in the vertical plane of QS #1 or horizontal plane of QS #2. Intuitively for a simple parabola, if the data sampled do not characterize the vertex, there are more possible reconstructions within the error of the data. This limitation in one plane of each quadrupole scan may introduce an error to those measurements that are not fully described by the statistical errors in Table 1. Relative errors are smaller or equal in the data that characterize the vertex, but not to a significant extent (relative statistical errors are 1-6% for all parameters).

A good covering of the vertex (i.e. transition of the scan through minimum of the measured beam size) in a simple quadrupole scan correlates with variation of the phase advance between the quadrupole and WS, which is conveniently viewed in normalized coordinates. Indeed, in Fig. 3 QS #1 characterizes 118º and 39º in X and Y, respectively, and QS #2 characterises 65º and 148º, respectively. The range of quadrupole currents in both scans are similar, and this change in the phase advance range between QS #1 and #2 originate from the adjustment to the upstream optics and thus initial conditions.

Expanding the quadrupole strengths measured will not improve the reconstruction. At the extreme quadrupole cur-rent, the beam transmission is lower, <95% marked with a lighter color in Fig. 1, and sizes deviate from the parabolic fit of the data. Lowered transmission implies the beam object has changed during transport. These data are unsuitable for reconstruction and were excluded from fitting.

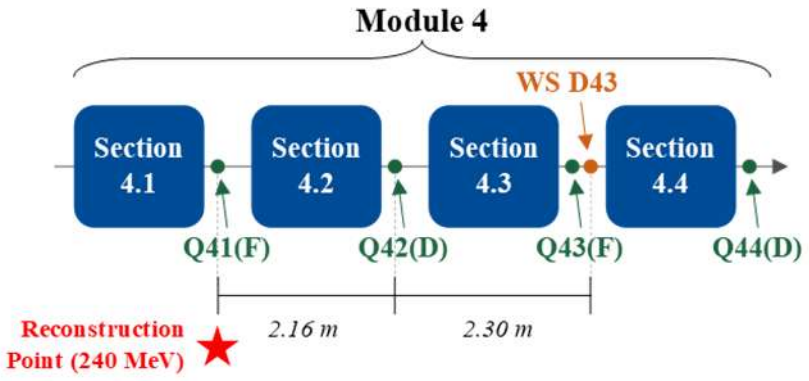

Figure 4: Module 4 quadrupole scan measurement design.

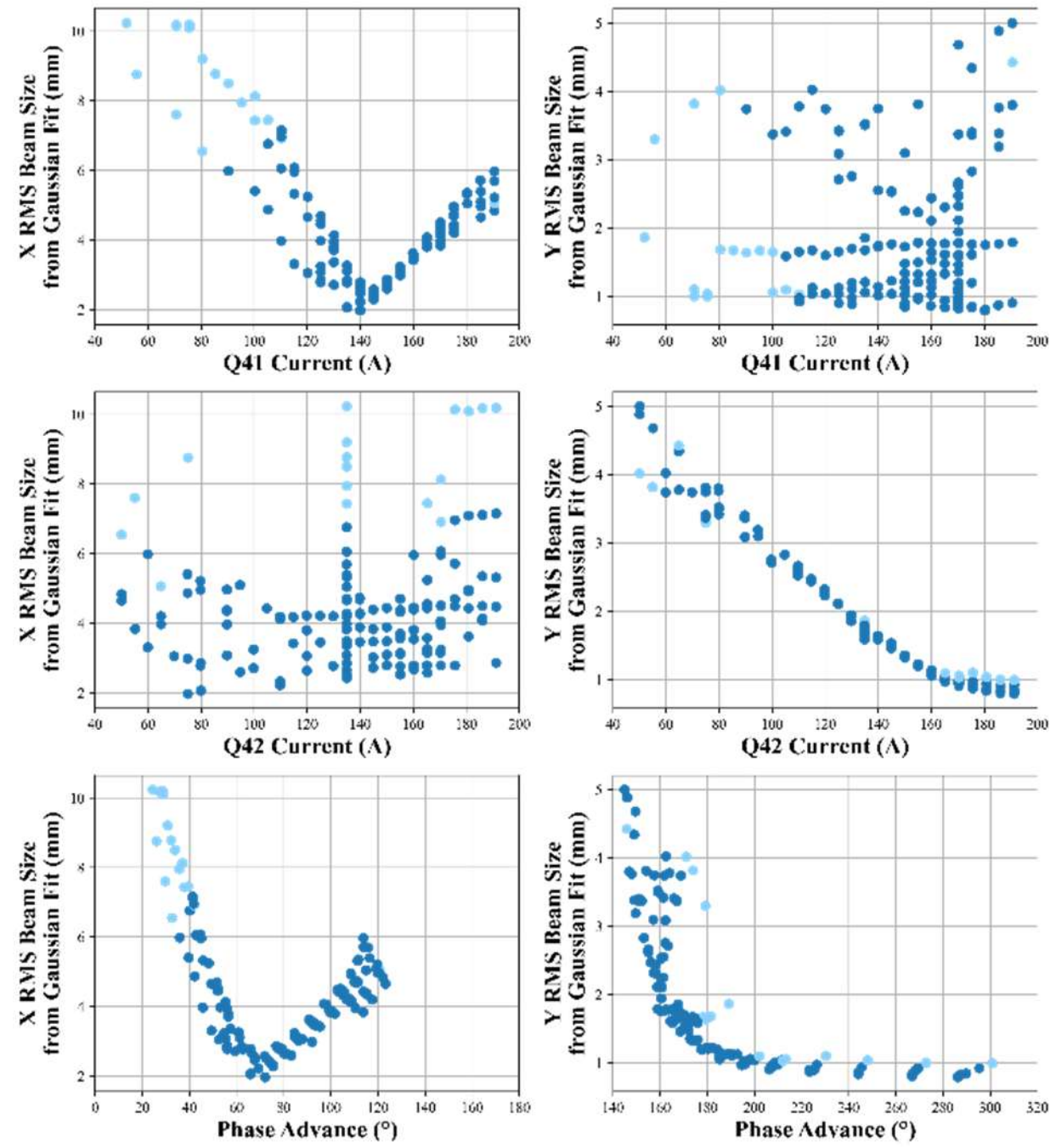

Figure 5: Measured beam size in X (left) and Y (right) with respect to Q41 current (top), Q42 current (middle), and simulated predictions for phase advance (bottom) in the Module 4 quadrupole scan.

## GENERALIZED QUADRUPOLE SCAN

The Module 4 location, Fig. 4, was selected for a generalized quadrupole scan using WS D43 and two driving quadrupoles, Q41 and Q42. Accelerating cavities in Module 4 were left on and the beamline was left under nominal conditions upstream. The reconstruction point is at Q41.

In a generalized measurement, there is no guidance as simple as a parabolic minimum. Rather than sampling the full range of Q41 and Q42 evenly, the currents selected for the measurements prioritize a wide range of phase advances and high transmission, as suggested by experience with the simple quadrupole scans. Simulations of Q41 and Q42 were used to make these selections. Simulated phase advance had a maximum gap of ~93º and ~30º in X and Y, respectively.

Data with transmission >95% were used for the reconstruction, "QS #3", and the results are in Table 1. Errors are calculated as described in Eq. (7). We expect emittance to be preserved between the 116 MeV and 240 MeV locations in the SCL, and the agreement is within the 3σ between QS #3 and QS #1, in the horizontal plane, and within 2σ between QS #3 and QS #2, in the vertical plane. Visualization of the beam size w.r.t. phase advance provides a clear description of how the beam object is sampled, Fig. 5.

Reconstruction from the simulated data informed the use of 95% as the transmission requirement for data inclusion, Fig. 6. In the horizontal plane, the reconstruction monotonically improves with strictness of the transmission; in the vertical plane, the reconstruction is already very good (≤11%) and varies around the initial conditions. The variation in the vertical plane may be attributed to losses originating from large horizontal beam sizes, Fig. 5. Aperture limitations suggest beam must remain below 7.5 mm (15 mm radius aperture, 2σ = 95%) at worst.

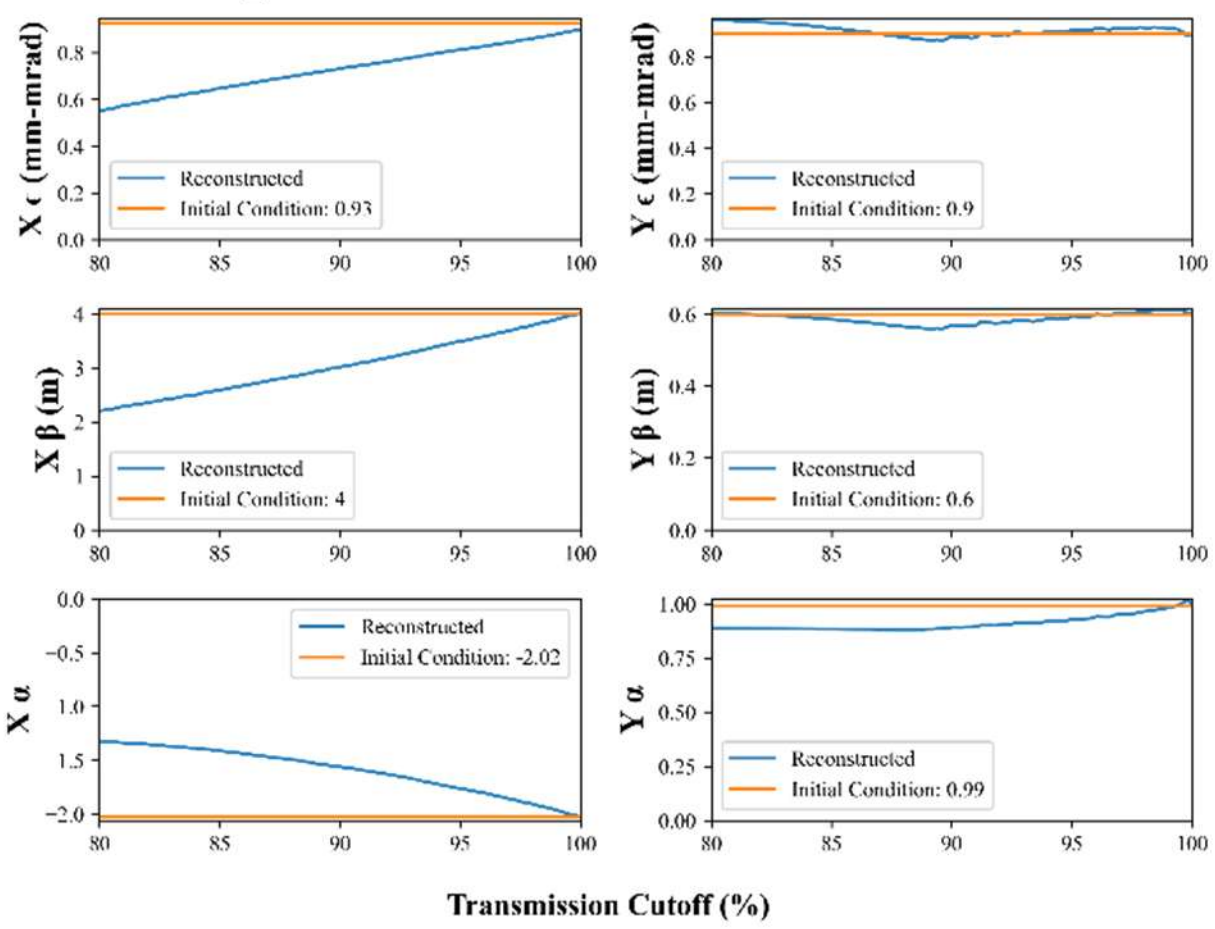

Figure 6: Changes in reconstructed values from simulations of QS #3 under varying transmission cutoffs.

## SUMMARY

The beam second moments are measured in quadrupole scan measurements at two locations in Fermilab Side-Coupled Linac, supporting development of an optics model for beam tuning [8] and preparations for the Linac2 commissioning [9]. The experience highlights the need for designing a quadrupole scan in advance, particularly to provide a large variation in the phase advance for efficient sampling of the beam and low beam losses between the detector and reconstruction point. Visualization in normalized coordinates facilitates such preparation.